\documentclass[letterpaper]{article} 
\usepackage{aaai24}  
\usepackage{times}  
\usepackage{helvet}  
\usepackage{courier}  
\usepackage[hyphens]{url}  
\usepackage{graphicx} 
\usepackage{natbib}  
\usepackage{caption} 
\usepackage{algorithm}
\usepackage{algorithmic}
\usepackage{enumitem}
\usepackage{booktabs}
\usepackage{amsmath}
\usepackage[dvipsnames]{xcolor}
\usepackage{mdframed}
\usepackage{makecell}
\usepackage{subcaption}

\usepackage{newfloat}
\usepackage{listings}
\DeclareCaptionStyle{ruled}{labelfont=normalfont,labelsep=colon,strut=off} 
\floatstyle{ruled}
\newfloat{listing}{tb}{lst}{}
\floatname{listing}{Listing}
\title{Recommender AI Agent: Integrating Large Language Models for Interactive Recommendations}
\author{
    Xu Huang\textsuperscript{\rm 1}, 
    Jianxun Lian\textsuperscript{\rm 2}\footnote{Corresponding authors.},
    Yuxuan Lei\textsuperscript{\rm 1}, 
    Jing Yao\textsuperscript{\rm 2}, 
    Defu Lian\textsuperscript{\rm 1}\footnotemark[1], 
    Xing Xie\textsuperscript{\rm 2}
}
\affiliations{
    \textsuperscript{\rm 1}School of Computer Science and Technology, University of Science and Technology of China, Hefei, China\\
    \textsuperscript{\rm 2}Microsoft Research Asia, Beijing China\\

    xuhuangcs@mail.ustc.edu.cn, jianxun.lian@outlook.com, lyx180812@mail.ustc.edu.cn, \\
    jingyao@microsoft.com,  liandefu@ustc.edu.cn, xingx@microsoft.com
}

\usepackage{bibentry}

\begin{document}

\nocopyright
\maketitle

\begin{abstract}


Recommender models excel at providing domain-specific item recommendations by leveraging extensive user behavior data. Despite their ability to act as lightweight domain experts, they struggle to perform versatile tasks such as providing explanations and engaging in conversations. On the other hand, large language models (LLMs) represent a significant step towards artificial general intelligence, showcasing remarkable capabilities in instruction comprehension, commonsense reasoning, and human interaction. However, LLMs lack the knowledge of domain-specific item catalogs and behavioral patterns, particularly in areas that diverge from general world knowledge, such as online e-commerce. Finetuning LLMs for each domain is neither economic nor efficient. 

In this paper, we bridge the gap between recommender models and LLMs, combining their respective strengths to create a versatile and interactive recommender system. We introduce an efficient framework called \textbf{InteRecAgent}, which employs LLMs as the brain and recommender models as tools. We first outline a minimal set of essential tools required to transform LLMs into InteRecAgent. We then propose an efficient workflow within InteRecAgent for task execution, incorporating key components such as memory components, dynamic demonstration-augmented task planning, and reflection. InteRecAgent enables traditional recommender systems, such as those ID-based matrix factorization models, to become interactive systems with a natural language interface through the integration of LLMs. Experimental results on several public datasets show that InteRecAgent achieves satisfying performance as a conversational recommender system, outperforming general-purpose LLMs. 
The source code of InteRecAgent is released at \url{https://aka.ms/recagent}.

\end{abstract}

\section{Introduction}

Recommender systems (RSs) have become an essential component of the digital landscape, playing a significant role in helping users navigate the vast array of choices available across various domains such as e-commerce and entertainment. By analyzing user preferences, historical data, and contextual information, these systems can deliver personalized recommendations that cater to individual tastes. Over the years, recommender systems have evolved from simple collaborative filtering algorithms to more advanced hybrid approaches that integrate deep learning techniques. However, as users increasingly rely on conversational interfaces for discovering and exploring products, there is a growing need to develop more sophisticated and interactive recommendation systems that can understand and respond effectively to diverse user inquiries and intents in an conversational manner.

Large language models (LLMs), such as GPT-3~\cite{brown2020language} and PaLM~\cite{chowdhery2022palm}, have made significant strides in recent years, demonstrating remarkable capabilities in artificial general intelligence and revolutionizing the field of natural language processing. A variety of practical tasks can be accomplished in the manner of users conversing with AI agents such as ChatGPT~\footnote{\url{https://chat.openai.com/}} and Claude~\footnote{\url{https://claude.ai/}}.
With their ability to understand context, generate human-like text, and perform complex reasoning tasks, LLMs can facilitate more engaging and intuitive interactions between users and RSs, thus offering promising prospects for the next generation of RSs. By integrating LLMs into RSs, it becomes possible to provide a more natural and seamless user experience that goes beyond traditional recommendation techniques, fostering a more timely understanding of user preferences and delivering more comprehensive and  persuasive suggestions.

Despite their potential, leveraging LLMs for recommender systems is not without its challenges and limitations. Firstly, while LLMs are pretrained on vast amounts of textual data from the internet, covering various domains and demonstrating impressive general world knowledge, they may fail to capture fine-grained, domain-specific behavior patterns, especially in domains with massive training data. Secondly, LLMs may struggle to understand a domain well if the domain data is private and less openly accessible on the internet. Thirdly, LLMs lack knowledge of new items released after the collection of pretraining data, and fine-tuning with up-to-date data can be prohibitively expensive. In contrast, in-domain models can naturally address these challenges. A common paradigm to overcome these limitations is to combine LLMs with in-domain models, thereby filling the gaps and producing more powerful intelligence. Notable examples include AutoGPT~\footnote{\url{https://github.com/Significant-Gravitas/Auto-GPT}}, HuggingGPT\cite{shen2023hugginggpt}, and Visual ChatGPT\cite{wu2023visualchatgpt}. The core idea is to utilize LLMs as the ``brains'' and in-domain models as ``tools'' that extend LLMs' capabilities when handling domain-specific tasks. 

In this paper, we connect LLMs with traditional recommendation models for interactive recommender systems. We propose InteRecAgent (\textbf{Inte}ractive \textbf{Rec}ommender \textbf{Agent}), a framework explicitly designed to cater to the specific requirements and nuances of recommender systems, thereby establishing a more effective connection between the LLM's general capabilities and the specialized needs of the recommendation domain. 
This framework consists of three distinct sets of tools, including querying, retrieval, and ranking, which are designed to cater to the diverse needs of users' daily inquiries.
Given the typically large number of item candidates, storing item names in the tools' input and output as observations with prompts is impractical. Therefore, we introduce a ``shared candidate bus'' to store intermediate states and facilitate communication between tools.
To enhance the capabilities of dealing with long conversations and even lifelong conversations, we introduce a ``long-term and short-term user profile'' module to track the preferences and history of the user, leveraged as the input of the ranking tool to improve personalization. The ``shared candidate bus'' along with the ``long-term and short-term user profile'' constitute the advanced memory mechanisms within the InteRecAgent framework.

Regarding task planning, we employ a ``plan-first execution'' strategy as opposed to a \textsl{step-by-step} approach. This strategy not only lowers the inference costs of LLMs but can also be seamlessly integrated with the dynamic demonstration strategy to enhance the quality of plan generation. Specifically, InteRecAgent generates all the steps of tool-calling at once and strictly follows the execution plan to accomplish the task. During the conversation, InteRecAgent parses the user's intent and retrieves a few demonstrations that are most similar to the current intent. These dynamically retrieved demonstrations help LLMs formulate a correct task execution plan. 
In addition, we implement a reflection strategy, wherein another LLM acts as a critic to evaluate the quality of the results and identify any errors during the task execution. If the results are unsatisfactory or errors are detected, InteRecAgent reverts to the initial state and repeats the plan-then-tool-execution process. 

Employing GPT-4 as the LLM within InteRecAgent has yielded impressive results in our experiments. This naturally leads to the attractive question: is it possible to harness a smaller language model to act as the brain? To explore this, we have developed an imitation dataset featuring tool plan generations derived from interactions between InteRecAgent and a user simulator, both powered by GPT-4. Through fine-tuning the LlaMA 2~\cite{touvron2023llama2} model with this dataset, we have created RecLlama. Remarkably, RecLlama surpasses several larger models in its effectiveness as the core of a recommender agent.
Our main contributions are summarized as follows:
\begin{itemize}[leftmargin=*]
\item We propose InteRecAgent, a compact LLM-based agent framework that democratizes interactive recommender systems by connecting LLMs with three distinct sets of traditional recommendation tools.   

\item In response to the challenges posed by the application of LLM-based agents in recommendation systems, we introduce a suite of advanced modules, including shared candidate bus, long-term and short-term user profile, dynamic demonstration-augmented plan-first strategy, and a reflection strategy.

\item To enable small language models to serve as the brain for recommender agents,  we create an imitation dataset derived from GPT-4. Leveraging this dataset, we have successfully fine-tuned a 7-billion-parameter model, which we refer to as RecLlama.

\item Experimental results from three public datasets demonstrate the effectiveness of InteRecAgent, with particularly significant advantages in domains that are less covered by world knowledge. 

\end{itemize}

\section{Related Work}

\subsection{Conversational Recommender System}
Existing researches in conversational recommender systems (CRS) can be primarily categorized into two main areas~\cite{gao2021advances}: attribute-based question-answering\cite{zou2019learning,zou2020towards,xu2021adapting} and open-ended conversation~\cite{li2018towards,wang2022towards,wang2021recindial}.
In attribute-based question-answering CRS, the system aims to recommend suitable items to users within as few rounds as possible. The interaction between the system and users primarily revolves around question-answering concerning desired item attributes, iteratively refining user interests. Key research challenges in this area include developing strategies for selecting queried attributes\cite{mirzadeh2005feature,zhang2018towards} and addressing the exploration-exploitation trade-off\cite{christakopoulou2016towards,xie2021comparison}.
In open-ended conversation CRS, the system manages free-format conversational data. Initial research efforts in this area focused on leveraging pretrained language models for conversation understanding and response generation\cite{li2018towards,penha2020does}. Subsequent studies incorporated external knowledge to enhance the performance of open-ended CRS\cite{chen2019towards,wang2022barcor,wang2022towards}. Nevertheless, these approaches struggle to reason with complex user inquiries and maintain seamless communication with users. The emergence of LLMs presents an opportunity to revolutionize the construction of conversational recommender systems, potentially addressing the limitations of existing approaches and enhancing the overall user experience.
   

\subsection{Enhancing LLMs}

The scaling-up of parameters and data has led to significant advancements in the capabilities of LLMs, including in-context learning~\cite{brown2020language,liu2021makes,rubin2021learning}, instruction following~\cite{ouyang2022training,touvron2023llama,openai2023gpt4}, planning and reasoning~\cite{wei2022chain,wang2022self,yao2022react,yang2023mm,wang2023plan}. In recommender systems, the application of LLMs is becoming a rapidly growing trend~\cite{liu2023chatgpt,dai2023uncovering,kang2023llms,wang2023zero}.


As models show emergent intelligence, researchers have started exploring the potential to leverage LLMs as autonomous agents~\cite{wang2023survey,zhao2023depth}, augmented with memory modules, planning ability, and tool-using capabilities. For example, \cite{wang2023augmenting,zhong2023memorybank,liu2023think} have equipped LLMs with an external memory, empowering LLMs with growth potential. Regarding the planning, CoT~\cite{wei2022chain,kojima2022large} and ReAct~\cite{yao2022react} propose to enhance planning by step-wise reasoning; ToT~\cite{yao2023tree} and GoT~\cite{besta2023graph} introduce multi-path reasoning to ensure consistency and correctness; Self-Refine~\cite{madaan2023self} and Reflexion~\cite{shinn2023reflexion} lead the LLMs to reflect on errors, with the ultimate goal of improving their subsequent problem-solving success rates. To possess domain-specific skills, some works~\cite{qin2023tool} study guiding LLMs to use external tools, such as a web search engine~\cite{nakano2021webgpt,shuster2022blenderbot}, mathematical tools~\cite{schick2023toolformer,thoppilan2022lamda}, code interpreters~\cite{gao2023pal,chen2022program} and visual models~\cite{wu2023visualchatgpt,shen2023hugginggpt}. To the best of our knowledge, this paper is the first to explore the LLM + tools paradigm in the field of recommender systems.

\section{Methodologies}
\begin{figure*}[htb]
    \centering
    \includegraphics[width=2.0\columnwidth]{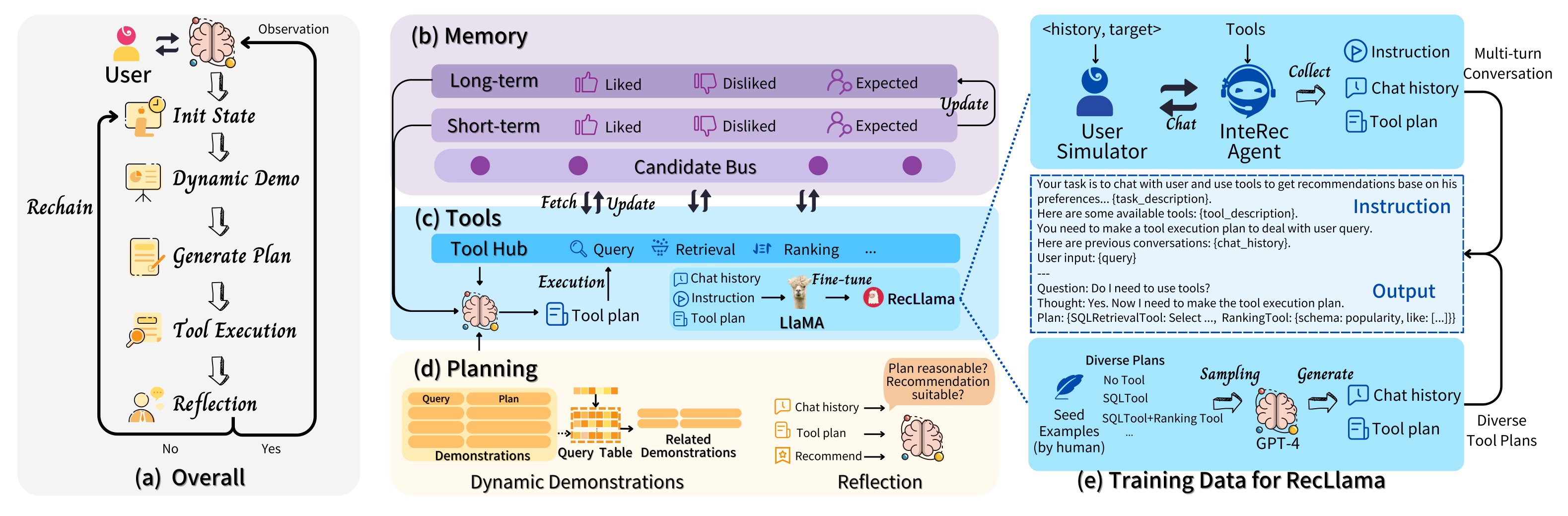}
    \vspace{-0.3cm}
    \caption{InteRecAgent Framework. (a) The overall pipeline of InteRecAgent; (b) The memory module, consisting of a candidate memory bus, a long-term and a short-term user profile; (c) Tool module, consisting of various tools, the plan-first execution strategy and the fine-tuning of RecLlama; (d) Planning module, involving the dynamic demonstrations and the reflection strategy; (e) Sources of fine-tuning data for RecLlama.}
    \label{fig: framework}
\end{figure*}


\subsection{The Overall Framework}

The comprehensive framework of InteRecAgent is depicted in Figure~\ref{fig: framework}. Fundamentally, LLMs function as the brain, while recommendation models serve as tools that supply domain-specific knowledge. Users engage with an LLM using natural language. The LLM interprets users' intentions and determines whether the current conversation necessitates the assistance of tools. For instance, in a casual chit-chat, the LLM will respond based on its own knowledge; whereas for in-domain recommendations, the LLM initiates a chain of tool calls and subsequently generates a response by observing the execution results of the tools. Consequently, the quality of recommendations relies heavily on the tools, making the composition of tools a critical factor in overall performance. To ensure seamless communication between users and InteRecAgent, covering both casual conversation and item recommendations, we propose a minimum set of tools that encompass the following aspects:

   \textbf{(1) Information Query.} 
    During conversations, the InteRecAgent not only handles item recommendation tasks but also frequently addresses users' inquiries. For example, within a gaming platform, users may ask questions like, \textsl{``What is the release date of this game and how much does it cost?''} To accommodate such queries, we include an item information query module. This module can retrieve detailed item information from the backend database using Structured Query Language (SQL) expressions.
    
     \textbf{(2) Item Retrieval.} 
    Retrieval tools aim to propose a list of item candidates that satisfy a user's demand from the entire item pool. These tools can be compared to the retrieval stage of a recommender system, which narrows down relevant candidates to a smaller list for large-scale serving. In InteRecAgent, we consider two types of demands that a user may express in their intent: hard conditions and soft conditions. Hard conditions refer to explicit demands on items, such as \textsl{``I want some popular sports games''} or ``\textsl{Recommend me some RPG games under \$100''}. Soft conditions pertain to demands that cannot be explicitly expressed with discrete attributes and require the use of semantic matching models, like \textsl{``I want some games similar to Call of Duty and Fortnite''}. It is essential to incorporate multiple tools to address both conditions. Consequently, we utilize an SQL tool to handle hard conditions, finding candidates from the item database. For soft conditions, we employ an item-to-item tool that matches similar items based on latent embeddings.
    
    \textbf{(3) Item Ranking.} 
Ranking tools execute a more sophisticated prediction of user preferences on the chosen candidates by leveraging user profiles. Similar to the rankers in conventional recommender systems, these tools typically employ a one-tower architecture. The selection of candidates could emerge from the output of item retrieval tools or be directly supplied by users, as in queries like ``\textsl{Which one is more suitable for me, item A or item B?}''. Ranking tools guarantee that the recommended items are not only pertinent to the user's immediate intent but also consonant with their broader preferences.

LLMs have the potential to handle various user inquiries when supplemented with these diverse tools. For instance, a user may ask, ``\textsl{I've played Fortnite and Call of Duty before. Now, I want to play some puzzle games with a release date after Fortnite's. Do you have any recommendations?}'' In this scenario, the tool execution sequence would be ``SQL Query Tool $\to$ SQL Retrieval Tool $\to$ Ranker Tool.'' First, the release date of Fortnite is queried, then the release date and puzzle genre are interpreted as hard conditions for the SQL retrieval. Finally, \textsl{Fortnite} and \textsl{Call of Duty} are considered as the user profile for the ranking model.

Typically, the tool augmentation is implemented via ReAct~\cite{yao2022react}, where LLMs generate reasoning traces, actions, and observations in an interleaved manner. We refer to this style of execution as \textit{step-by-step}. Our initial implementation also employed the step-by-step approach. However, we soon observed some limitations due to various challenges. Firstly, retrieval tools may return a large number of items, resulting in an excessively long observation prompt for LLMs. Additionally, including numerous entity names in the prompt can degrade LLMs performance. Secondly, despite their powerful intelligence, LLMs may use tools incorrectly to complete tasks, such as selecting the wrong tool to call or omitting key execution steps. To tackle these challenges, we enhance the three critical components of a typical LLM-based agent, namely memory (Section~\ref{sec:memory}), task planning (Section~\ref{sec:planfirst} and  ~\ref{sec: reflection}), and tool learning abilities (Section~\ref{sec:recllama}).

\subsection{Memory Mechanism}\label{sec:memory}
\subsubsection{Candidate Bus}
The large number of items can pose a challenge when attempting to include items generated by tools in prompts as observations for the LLM, due to input context length limitations. Meanwhile, the input of a subsequent tool often depends on the output of preceding tools, necessitating effective communication between tools. Thus, we propose Candidate Bus, which is a separate memory to store the current item candidates, eliminating the need to append them to prompt inputs. The Candidate Bus, accessible by all tools, comprises two parts: a data bus for storing candidate items, and a tracker for recording each tool's output. 

The candidate items in the data bus are initialized to include all items at the beginning of each conversation turn by default. At the start of each tool execution, candidate items are read from the data bus, and the data bus is then refreshed with the filtered items at the end of each tool execution. This mechanism allows candidate items to flow sequentially through the various tools in a streaming manner. Notably, users may explicitly specify a set of candidate items in the conversation, such as ``\textsl{Which of these movies do you think is most suitable for me: [Movie List]?}'' In this case, the LLM will call a special tool—the memory initialization tool—to set the user-specified items as the initial candidate items.

The tracker within the memory serves to record tool execution. Each tool call record is represented as a triplet $(f_k, i_{k}, o_{k})$, where $f_k$ denotes the name of the $k$-th tool, and $i_{k}, o_{k}$ are the input and output of the tool's execution, such as the number of remaining candidates, runtime errors. The tracker's main function is to aid the critic in making judgments within the reflection mechanism, acting as the $\boldsymbol{o}^t$ in $\operatorname{reflect}(\cdot)$, as described in Section~\ref{sec: reflection}.

With the help of the Candidate Bus component, items can be transmitted in a streaming manner between various tools and continuously filtered according to conditions, presenting a funnel-like structure for the recommendation. The tracker's records can be considered as short-term memory for further reflection. We depict an example of the memory bus in the upper of Figure~\ref{fig:example_review2}.

\subsubsection{User Profile}\label{sec:long-term}
To facilitate the invocation of tools, we explicitly maintain a user profile in memory. This profile is structured as a dictionary that encapsulates three facets of user preference: ``like'', ``dislike'', and ``expect''.  The ``like'' and ``dislike'' facets reflect the user's favorable and unfavorable tastes, respectively, whereas ``expect'' monitors the user's immediate requests during the current dialogue, such as conducting a search, which is not necessarily indicative of the user's inherent preferences. Each facet may contain content that includes item names or categories.

User profiles are synthesized by LLMs based on conversation history. To address situations where the conversation history grows excessively long, such as in lifelong learning scenarios where conversations from all days may be stored for ongoing interactions, we devise two distinct user profiles: one representing long-term memory and another for short-term memory. Should the current dialogue exceed the LLM's input window size, we partition the dialogue, retrieve the user profile from the preceding segment, and merge it with the existing long-term memory to update the memory state. The short-term memory is consistently derived from the most recent conversations within the current prompt. When it comes to tool invocation, a comprehensive user profile is formed by the combination of both long-term and short-term memories.

\subsection{Plan-first Execution with Dynamic Demonstrations}\label{sec:planfirst}
Rather than using the \textit{step-by-step} approach, we adopt a two-phase method. In the first phase, we prompt the LLM to generate a complete tool execution plan based on the user's intention derived from the dialogue. In the second phase, the LLM strictly adheres to the plan, calling tools in sequence while allowing them to communicate via the Candidate Bus. 
Concretely, the plan-first execution consists of the following two phases.

\begin{itemize}[leftmargin=*]
    \item {\textbf{Plan}}: LLM accepts the user's current input $x^t$, dialogue context $C^{t-1}$, descriptions of various tools $\mathcal{F}$, and demonstration $\mathcal{D}_{x^t}$ for in-context learning. LLM formulates a tool usage plan based on user intent and preferences, providing inputs for each tool, i.e., $\boldsymbol{p}^t = \{p^t_1,\cdots,p^t_n\} = \operatorname{plan}\left(x^t, C^{t-1}, \mathcal{F}, \mathcal{D}_{x^t}\right)$, where $p^t_k=(f_k,i_k)$ consists of the tool $f_k$ and its input $i_k$.
    \item {\textbf{Execution}}: The tool executor invokes the tools step-by-step according to the plan $\boldsymbol{p}_{t}$ and obtains outputs from each tool, i.e., $\boldsymbol{o}^{t}=\{o^t_1,\cdots,o^t_n\} =\operatorname{exec}(\boldsymbol{p}^t, \mathcal{F})$. The output feedback of each tool $f_k$ is defined as $o^t_k$, where only the item information $o^t_n$ from the last tool's output serves as LLM's observation for generating the response $y^t$. The remaining information is tracked by the candidate memory bus for further reflection (see Section~\ref{sec: reflection}).  
\end{itemize}

We summarize the differences between our plan-first execution strategy and \textsl{step-by-step} strategy in Table~\ref{tab: react_vs_plan-first} from six aspects. Fundamentally, \textsl{step-by-step} strategy executes reasoning and action execution alternately, while our plan-first execution is a two-phase strategy, where a series of executions is conducted followed by one-time planning. 
In \textsl{step-by-step} strategy, the LLMs are responsible for thinking and reasoning at each step. The task entails reasoning for individual observation, resulting in-context learning being challenging due to the difficulty in crafting demonstrations comprising dynamic observations. Differently, the primary task of LLM in our plan-first execution is to make a tool utilizing plan, which could be easily guided by $\langle\text{query},\text{plan}\rangle$ pairs. 
The foremost advantage of our plan-first execution resides in the reduction of API calls. When employing N steps to address a task, our strategy necessitates merely 2 API calls, as opposed to N+1 calls in ReAct. This leads to a decrease in latency, which is of particular importance in conversational settings.

\begin{table}[h]
    \centering
    \caption{Property Comparisons between ReAct and Plan-first Execution. ICL is the abbreviation of In-Context Learning. }
    \vspace{-0.3cm}
    \begin{tabular}{c|c|c} 
    \toprule
     Property & ReAct & Plan-first Exe \\ \midrule
     Basic Idea & step-wise reason & task-wise plan \\ \midrule
     ICL & hard & easy \\ \midrule
     Reflection & internal & external \\ \midrule
     \# API Call & N+1 & 2 \\ \midrule
     Latency & $(N+1)\Delta t_{api}+\Delta t_{exe}$  & $2\Delta t_{api}+\Delta t_{exe}$ \\
     \bottomrule
    \end{tabular}
    \label{tab: react_vs_plan-first}
\end{table}

In order to improve the planning capability of LLM, demonstrations $\mathcal{D}_{x^t}$ are injected into prompts for in-context learning in the \textbf{Plan} phase. Each demonstration consists of a user intent $x$ and tool execution path $\boldsymbol{p}$. However, the number of demonstrations is strictly limited by the contextual length that LLM can process, which makes the quality of demonstrations of paramount importance. To address the challenge, we introduce a dynamic demonstration strategy, where only a few demonstrations that are most similar to current user intent are incorporated into the prompt. For example, if the current user input is ``\textsl{My game history is Call of Duty and Fortnite, please give me some recommendations}'', then demonstration with user intent ``\textsl{I enjoyed} \texttt{ITEM1}\textsl{,} \texttt{ITEM2} \textsl{in the past, give me some suggestions}'' may be retrieved as a high-quality demonstration. 

Inspired by Self-Instruct~\cite{madaan2023self}, we use LLM to generate demonstrations of tool-using plans in the form of $(x, \boldsymbol{p})$. First, we manually write some (\string~20) typical user intents and the corresponding execution as seed demonstrations; then, we use the input-first and output-first strategies to generate more demonstrations using LLM. In the input-first strategy, there are two stages: first, the LLM generates $x$ by emulating the intents in seed demonstrations, and then the LLM makes plans $ \boldsymbol{p}$ for these intents. The output-first method consists of three stages: first, we provide the LLM with a plan $\boldsymbol{p}$ and generate corresponding user intent $x$. Then, we use LLM to make plans $\boldsymbol{\tilde{p}}$ for the intent, and finally, we verify whether the generated plan $\boldsymbol{\tilde{p}}$ is consistent with the given plan $\boldsymbol{p}$. The inconsistency indicates that the quality of the generated intent is not high enough, and we only retain those consistent demonstrations. The output-first method allows us to obtain demonstrations corresponding to all available plans, providing diversity for the demonstrations. Examples generated by input-first and output-first are illustrated in Figure~\ref{fig: example_demo}.

\subsection{Reflection} \label{sec: reflection}
Despite LLM's strong intelligence,  it still exhibits occasional errors in reasoning and tool utilization~\cite{madaan2023self, shinn2023reflexion}.
For example, it may violate instructions in the prompt by selecting a non-existent tool, omit or overuse some tools, or fail to prepare tool inputs in the proper format, resulting in errors in tool execution.

To reduce the occurrence of such errors, some studies have employed self-reflection~\cite{shinn2023reflexion} mechanisms to enable LLM to have some error-correcting capabilities during decision-making. In InteRecAgent, we utilize an \textbf{actor-critic} reflection mechanism to enhance the agent's robustness and the error-correcting ability. In the following part, we will formalize this self-reflection mechanism.

Assume that in the $t$-th round, the dialogue context is $C^{t-1}$ and the current user input is $x^t$. The actor is an LLM equipped with tools and inspired by the dynamic demonstration-augmented plan-first execution mechanism. For the user input, the actor would make a plan $\boldsymbol{p}^t$, obtain the tools' output $\boldsymbol{o}^t$ and generate the response $y^t$.
The critic evaluates the behavioral decisions of the actor. The execution steps of the reflection mechanism are listed as follows:

\begin{figure}[thb]
    \centering
    \small
    \begin{mdframed}[backgroundcolor=gray!10]
        \textbf{Intent(by GPT-4)}: Can you suggest some TYPE1 and TYPE2 items based on my preferences: ITEM1, ITEM2, and ITEM3? \\
        \textbf{Plan(by GPT-4)}: 1. SQL Retrieval Tool (TYPE1 and TYPE2); 2. Ranking Tool (by preference using ITEM1, ITEM2, and ITEM3); 3. Candidate Fetching Tool. \\

        \textbf{Plan}: 1. Candidates Storing Tool (ITEM1, ITEM2, ITEM3); 2. SQL Retrieval Tool (TYPE); 3. ItemCF Retrieval Tool (ITEM); 4. Ranking Tool (by preference); 5. Candidate Fetching Tool.\\
        \textbf{Intent(by GPT-4)}: I have a list of items: ITEM1, ITEM2, ITEM3. I want a TYPE item that is similar to ITEM, and please rank them based on my preferences.
    \end{mdframed}
    \vspace{-0.3cm}
    \caption{Examples of generated demonstrations in game domain.}
    \label{fig: example_demo}
\end{figure}

\begin{figure}[htb]
    \centering
    \begin{subfigure}[b]{0.9\columnwidth}
        \centering
        \includegraphics[width=0.99\columnwidth]{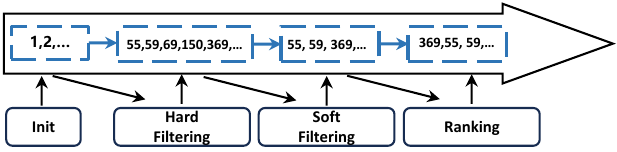}
        \vspace{0.1cm}
    \end{subfigure}
    \begin{subfigure}[b]{0.9\columnwidth}
        \centering
        \includegraphics[width=0.99\columnwidth]{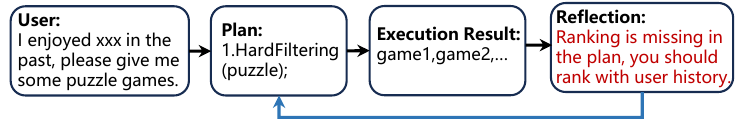}
    \end{subfigure}
    \vspace{-0.3cm}
    \caption{Example of memory bus (upper) and reflection (lower).}
    \label{fig:example_review2}
\end{figure}

\begin{itemize}[leftmargin=*]
\item \textbf{Step1}: The critic evaluates the actor's output $\boldsymbol{p}^t$, $\boldsymbol{o}^t$ and $y^t$ under the current dialogue context and obtains the judgment $\gamma = \operatorname{reflect}(x^t, C^{t-1}, \boldsymbol{p}^t, \boldsymbol{o}^t, y^t)$.
\item \textbf{Step2}: When the judgment $\gamma$ is positive, it indicates that the actor's execution and response are reasonable, and the response $y^t$ is directly provided to the user, ending the reflection phase. When the judgment $\gamma$ is negative, it indicates that the actor's execution or response is unreasonable. The feedback $\gamma$ is used as a signal to instruct the actor to rechain, which is used as the input of $\operatorname{plan}(\cdot)$. 
\end{itemize}

In the actor-critic reflection mechanism, the actor is responsible for the challenging plan-making task, while the critic is responsible for the relative simple evaluation task. The two agents cooperate on two different types of tasks and mutually reinforce each other through in-context interactions. This endows InteRecAgent with enhanced robustness to errors and improved error correction capabilities, culminating in more precise tool utilization and recommendations. An example of reflection is shown in the lower of Figure~\ref{fig:example_review2}.

\subsection{Tool Learning with Small Language Models}\label{sec:recllama}
The default LLM served as the brain is GPT-4, chosen for its exceptional ability to follow instructions compared to other LLMs. We are intrigued by the possibility of distilling GPT-4's proficiency in instruction-following to smaller language models (SLMs) such as the 7B-parameter Llama, aiming to reduce the costs associated with large-scale online services and to democratize our InteRecAgent framework to small and medium-sized business clients. To achieve this, we utilize GPT-4 to create a specialized dataset comprising pairs of [instructions, tool execution plans]. The ``instruction'' element encompasses both the system prompt and the user-agent conversation history, acting as the input to elicit a tool execution plan from the LLM; the ``tool execution plan'' is the output crafted by GPT-4, which serves as the target for fine-tuning Llama-7B. We denote the fine-tuned version of this model \textbf{RecLlama}.

To ensure the high quality of the RecLlama dataset, we employ two methods to generate data samples. The first method gathers samples from dialogues between a user simulator and a recommender agent, which is powered by GPT-4. Note that during one conversation, each exchange of user-agent produces one data sample, capturing the full range of GPT-4's responses to the evolving context of the conversation.  However, this method might not encompass a sufficiently diverse array of tool execution scenarios due to the finite number of training samples we can manage. Therefore, we complement this with a second method wherein we initially craft 30 varied dialogues designed to span a wide range of tool execution combinations. Then, for each iteration, we select three of these dialogues at random and prompt GPT-4 to generate both a conversation history and a suitable tool execution plan. This approach significantly enhances the diversity of the RecLlama dataset.

To evaluate RecLlama's capacity for domain generalization, we limit the generation of training data to the Steam and MovieLens datasets, excluding the Beauty dataset (the details of datasets will be elaborated in Section~\ref{sec:dataset}). The final RecLlama dataset comprises 16,183 samples, with 13,525 derived from the first method and 2,658 from the second.


\section{Experiments}


\subsection{Experimental Setup}

\subsubsection{Evaluation Strategies.}

Evaluating conversational recommender systems presents a challenge, as the seeker communicates their preferences and the recommendation agent provides suggestions through natural, open-ended dialogues. To enable the quantitative assessment of InteRecAgent, we design the following two evaluation strategies:

\textbf{(1) User Simulator.} We manually tune a role-playing prompt to facilitate GPT-4 in emulating real-world users with varying preferences. A simulated user's preference is ascertained by injecting their historical behaviors into the role-playing prompt, leaving out the last item in their history as the target of their next interest. Following this, the simulated user engages with the recommendation agent to discover content that fits their interest.  In this way, GPT-4 operates from the standpoint of the user, swiftly reacting to the recommended outcomes, thereby crafting a more natural dialogue scenario. This approach is utilized to assess the efficacy of InteRecAgent within \textbf{multi-turn dialogue} settings. An illustrative example of a user simulator prompt can be found in  Figure~\ref{fig: user_simulator}.

\begin{figure}[thb]
    \centering
    \small
    \begin{mdframed}[backgroundcolor=gray!20]
        You are a user chatting with a recommender for \{item\} recommendation in turn. 
        Your history is \{history\}. Your target items: \{target\}. 
        Here is the information about target you could use: \{target\_item\_info\}. \\
        You must follow the rules below during chat. \\
        If the recommender recommends \{target\}, you should accept. 
        If the recommender recommends other items, you should refuse them and provide the information about \{target\}. 
        If the recommender asks for your preference, you should provide the information about \{target\}. \\
        You could provide your history. 
        Your output is only allowed to be the words from the user you act. 
        If you think the conversation comes to an ending, output a $\langle \text{END}\rangle$. 
        You should never directly tell the target item. 
        Only use the provided information about the target.  
        Never give many details about the target items at one time. Less than 3 conditions is better. \\
        Now lets start, you first, act as a user.
        Here are the previous conversation you have completed: \{chat\_history\}.
    \end{mdframed}
    \vspace{-0.3cm}
    \caption{Prompt for user simulator.}
    \label{fig: user_simulator}
\end{figure}

The default configuration for the user simulator is set to ``\textbf{session-wise}''. This implies that the agent will only access content within the current dialogue session, and its memory will be cleared once the user either successfully locates what they are seeking or fails to do so. The conversation turns in ``session-wise'' setting is usually limited, thus, the long-term memory module in InteRecAgent will not be activated. In order to assess the performance while handling ``\textbf{lifelong memory}'' (refer to Section~\ref{sec:long-term}), we have formulated two strategies for simulating extended dialogues. The first strategy, referred to as \textsc{Long-Chat}, mandates extended conversations between the user and the recommendation agent. This is achieved by alternately incorporating three types of chat intents within the user simulator: sharing history, detailing the target item, and participating in casual conversation. The simulator alternates between providing information (either historical or target-related) and casual chat every five rounds. During this process, if the agent mentions the target item, the conversation can be terminated and labeled as a success. The second strategy, referred to as \textsc{Long-Context}, initially synthesizes multi-day conversations utilizing user history. Subsequently, based on these extended dialogues, the user simulator interacts with the agent in a manner akin to the ``session-wise'' setting. For our method, the lengthy conversation history is loaded into the long-term memory module. However, for baseline methods, the extended conversation history will be truncated if it surpasses the maximum window size of the LLM.

\textbf{(2) One-Turn Recommendation.} 
Following the settings of traditional conversational recommender systems on ReDial~\cite{li2018towards}, we also adopt the one-turn recommendation strategy. Given a user's history, we design a prompt that enables GPT-4 to generate a dialogue, thereby emulating the interaction between a user and a recommendation agent. The objective is to ascertain whether the recommendation agent can accurately suggest the ground truth item in its next response. We assess both the item retrieval task (retrieval from the entire space) and the ranking task (ranking of provided candidates). Specifically, the dialogue context is presented to the recommendation agent, accompanied by the instruction \textit{Please give me k recommendations based on the chat history} for the retrieval task, and the instruction \textit{Please rank these candidate items based on the chat history} for the ranking task. To ensure a fair comparison with baseline LLMs, the \textsl{One-Turn Recommendation} evaluation protocol employs only the ``session-wise'' setting, and the long-term memory module in InteRecAgent remains deactivated.


\subsubsection{Dataset.}\label{sec:dataset}
To compare methods across different domains, we conduct experiments using three datasets: Steam\footnote{\url{https://github.com/kang205/SASRec}}, MovieLens\footnote{\url{https://grouplens.org/datasets/movielens/10m}} and Amazon Beauty\footnote{\url{http://jmcauley.ucsd.edu/data/amazon/links.html}}. Each dataset comprises user-item interaction history data and item metadata. We apply the leave-one-out method to divide the interaction data into training, validation, and testing sets. The training of all utilized tools is performed on the training and validation sets. Due to budget constraints, we randomly sample 1000 and 500 instances from the testing set for user simulator and one-turn benchmarking respectively. For the lifelong simulator, due to the costly long conversation, we use 100 instances in evaluation.

\subsubsection{Baselines.}
As dialogue recommendation agents, we compare our methods with the following baselines:  
\begin{itemize}[leftmargin=*]
    \item \textbf{Random}: Sample k items uniformly from entire item set.
    \item \textbf{Popularity}: Sample k items with item popularity as the weight.
    \item \textbf{LlaMA-2-7B-chat}, \textbf{LlaMA-2-13B-chat}~\cite{touvron2023llama2}: The second version of the LlaMA model released by Meta.  
    \item \textbf{Vicuna-v1.5-7B}, \textbf{Vicuna-v1.5-13B}~\cite{vicuna2023}: Open-source models fine-tuned with user-shared data from the ShareGPT\footnote{\url{https://sharegpt.com/}} based on LlaMA-2 foundation models. 
    \item \textbf{Chat-Rec}~\cite{gao2023chatrec}: A recently proposed conversational recommendation agent utilizes a text-embedding tool (OpenAI text-embedding-ada-002) to retrieve candidates. It then processes the content with an LLM before responding to users. We denote the use of GPT-3.5 as the LLM in the second stage with "Chat-Rec (3.5)" and the use of GPT-4 with "Chat-Rec (4)".
    \item \textbf{GPT-3.5}, \textbf{GPT-4}~\cite{openai2023gpt4}: We access these LLMs from OpenAI by API service. The GPT-3.5 version in use is gpt-3.5-turbo-0613 and GPT-4 version is gpt-4-0613\footnote{\url{https://platform.openai.com/docs/models/}}.  
 
\end{itemize}

For the LlaMA and Vicuna models, we employ the FastChat~\cite{zheng2023judging} package to establish local APIs, ensuring their usage is consistent with GPT-3.5 and GPT-4.

\begin{table}[t]
\small
\centering
\caption{Performance comparisons with the user simulator strategy (session-wise). H@5 is an abbreviation for Hit@5.}
\setlength{\tabcolsep}{4pt} 
\begin{tabular}{@{}l|ll|ll|ll@{}}
\toprule
 & \multicolumn{2}{c|}{Steam} & \multicolumn{2}{c|}{MovieLens} & \multicolumn{2}{c}{Beauty} \\ \midrule
Methods & H@5$\uparrow$  & AT@5$\downarrow$ & H@5$\uparrow$ & AT@5$\downarrow$ & H@5$\uparrow$  & AT@5$\downarrow$  \\ \midrule
LlaMA2-7B & 0.36 & 4.76 & 0.50 & 4.71 & 0.03 & 5.91 \\
LlaMA2-13B & 0.39 & 4.56 & 0.53 & 4.52 & 0.05 & 5.87 \\
Vicuna-7B  & 0.38 & 4.70 & 0.51 & 4.70 & 0.03 & 5.90 \\ 
Vicuna-13B   & 0.40 & 4.60 & 0.54 & 4.56 & 0.07 & 5.85 \\\midrule
Chat-Rec(3.5)  &  0.74 & 3.63 & 0.76 & 3.78 & 0.39 & 4.89 \\
Chat-Rec(4)  &  \underline{0.83} & 3.42 & \underline{0.82}  & \underline{3.62} & \underline{0.40} & \underline{4.80} \\
GPT-3.5  &  0.69 & 3.68 & 0.75 & 3.75 & 0.13 & 5.68 \\
GPT-4    & 0.78 & \underline{3.34} & 0.79 & 3.70 & 0.15 & 5.59  \\ \midrule
Ours  &  \textbf{0.87}  &  \textbf{2.86}  & \textbf{0.85}   & \textbf{3.15}  & \textbf{0.54} & \textbf{3.99}   \\ \bottomrule
\end{tabular}
\label{tab: simulator}
\end{table}

\subsubsection{Metrics.}
Since both our method and baselines utilize LLMs to generate response, which exhibit state-of-the-art text generation capabilities, our experiments primarily compare recommendation performance of different methods.  
For the \textsl{user simulator} strategy, we employ two metrics: Hit@$k$ and AT@$k$, representing the success of recommending the target item within $k$ turns and the average turns (AT) required for a successful recommendation, respectively. Unsuccessful recommendations within k rounds are recorded as $k+1$ in calculating AT. In the \textsl{one-turn} strategy, we focus on the Recall@$k$ and NDCG@$k$ metric for retrieval and ranking task, respectively. 
In Recall@$k$, the $k$ represents the retrieval of $k$ items, whereas in NDCG@$k$, the $k$ denotes the number of candidates to be ranked.

\subsubsection{Implementation Details.}
We employ GPT-4 as the brain of the InteRecAgent for user intent parsing and tool planing. Regarding tools, we use SQL as information query tool, SQL and ItemCF~\cite{linden2003amazon} as hard condition and soft condition item retrieval tools, respectively, and SASRec~\cite{kang2018self} without position embedding as the ranking tool. SQL is implemented with SQLite integrated in pandasql\footnote{\url{https://github.com/yhat/pandasql/}} and retrieval and ranking models are implemented with PyTorch. The framework of InteRecAgent is implement with Python and LangChain\footnote{\url{https://www.langchain.com/}}. For dynamic demonstration selection, we employ sentence-transformers\footnote{\url{https://huggingface.co/sentence-transformers}} to encode demonstrations into vectors and store them using ChromaDB\footnote{\url{https://www.trychroma.com/}}, which facilitates ANN search during runtime.
Regarding hyperparameter settings, we set the number of dynamic demonstrations to 3, the maximum number of candidates for hard condition retrieval to 1000, and the threshold for soft condition retrieval cut to the top 5\%. 

\subsection{Evaluation with User Simulator} \label{sec: exp_simulator}
\subsubsection{Session-wise setting.}
Table~\ref{tab: simulator} presents the results of evaluations conducted using the user simulator strategy. Our method surpasses other LLMs in terms of both hit rate and average turns across the three datasets. These results suggest that our InteRecAgent is capable of delivering more accurate and efficient recommendations in conversations compared to general LLMs.  Overall, LLMs with larger parameter sizes perform better. GPT-3.5 and GPT4, with parameter sizes exceeding 100B, significantly outperform LlaMA2 and Vicuna-v1.5 13B models from the same series almost always surpass 7B models, except for LlaMA2-7B and LlaMA2-13B, which both perform extremely poorly on the Beauty dataset. 

Another interesting observation is the more significant improvement in relatively private domains, such as Amazon Beauty. In comparison to gaming and movie domains, the beauty product domain is more private, featuring a larger number of items not well-covered by common world knowledge or being new. Table~\ref{tab: simulator} reveals that GPT-3.5 and GPT-4 exhibit competitive performance in gaming and movie domains. 
However, in the Amazon Beauty domain, most LLMs suffer severe hallucination issue due to the professional, long, and complex item names, resulting in a significant drop in performance. This phenomenon highlights the necessity of recommender agents in private domains. Leveraging the text embedding retrieval tool, Chat-Rec shows superior performance compared to GPT-3.5 and GPT-4, but still falling short of the performance achieved by InteRecAgent. Chat-Rec can be seen as a simplified version of InteRecAgent,  incorporating just a single tool within the agent's framework. Consequently, Chat-Rec lacks the capability to handle multifaceted queries, such as procuring detailed information about an item or searching for items based on intricate criteria.

\begin{table}[t]
\centering
\small
\caption{Performance comparisons with the user simulator strategy(\textsc{Long-Chat}). "+LT Mem." means activating the long-term memory module in our InteRecAgent. The higher Hit@50 and the lower AT@50, the better performance.}
\vspace{-0.3cm}
\setlength{\tabcolsep}{4pt} 
\begin{tabular}{@{}l|ll|ll|ll@{}}
\toprule
 & \multicolumn{2}{c|}{Steam} & \multicolumn{2}{c|}{MovieLens} & \multicolumn{2}{c}{Beauty} \\ \midrule
Methods & H@50  & AT@50 & H@50 & AT@50 & H@50  & AT@50  \\ \midrule
GPT-4    & {0.70} & {20.56} & {0.71} & {24.06} & 0.06 & 49.42  \\ \midrule
Ours  &  {0.83}  &  \textbf{16.85}  & {0.76}   & {20.13}  & {0.69} & {27.14}   \\
\makecell[l]{+LT Mem.}  &  \textbf{0.86}  &  {17.58}  & \textbf{0.77}   & \textbf{20.06}  & \textbf{0.74} & \textbf{25.88}   \\ \bottomrule
\end{tabular}
\label{tab: lifelong-simulator}
\end{table}

\begin{table}[t]
\centering
\small
\caption{Performance comparisons with the lifelong user simulator strategy(\textsc{Long-Context}). "+LT Mem." means activating the long-term memory module in our InteRecAgent. }
\vspace{-0.3cm}
\setlength{\tabcolsep}{4pt} 
\begin{tabular}{@{}l|ll|ll|ll@{}}
\toprule
 & \multicolumn{2}{c|}{Steam} & \multicolumn{2}{c|}{MovieLens} & \multicolumn{2}{c}{Beauty} \\ \midrule
Methods & H@5$\uparrow$  & AT@5$\downarrow$ & H@5$\uparrow$ & AT@5$\downarrow$ & H@5$\uparrow$  & AT@5$\downarrow$  \\ \midrule
GPT-4    & {0.74} & {3.05} & {0.82} & {3.03} & {0.09}  & {5.71}  \\ \midrule
Ours  &  {0.76}  &  {2.92}  & \textbf{0.83}   & {3.29}  & {0.38} & {4.58}   \\
\makecell[l]{+LT Mem.}  &  \textbf{0.79}  &  \textbf{2.70}  & \textbf{0.83}   & \textbf{2.84}  & \textbf{0.51} & \textbf{3.99}   \\ \bottomrule
\end{tabular}
\label{tab: lifelong-simulator2}
\end{table}

\subsubsection{Lifelong conversation setting.}
Table~\ref{tab: lifelong-simulator} and Table~\ref{tab: lifelong-simulator2} demonstrate the performance of two lifelong memory configurations, specifically, \textsc{Long-Chat} and \textsc{Long-Context}. For \textsc{Long-Chat}, the recommender agent engages a maximum of 50 rounds of dialogue with the user simulator. In both configurations, InteRecAgent without long-term memory modules (denoted as ``Ours'' in the tables) consistently outperforms GPT-4 across all datasets, which validates the robustness of our tool-enhanced recommender agent framework. After activating the long-term memory modules, the performance gets further improved under both \textsc{Long-Chat} and \textsc{Long-Context} configurations. This confirms the necessity and effectiveness of memory on capturing user preference during lifelong interactions between the user and AI agent.


\subsection{Evaluation with One-Turn Recommendation}
\begin{table}[t]
\centering
\small
\caption{Performance comparisons in one-turn recommendation (\%). R@5 and N@20 are abbreviations for Recall@5 and NDCG@20 respectively.}
\vspace{-0.3cm}
\setlength{\tabcolsep}{4.5pt} 
\begin{tabular}{@{}l|lll|lll@{}}
\toprule
 Task & \multicolumn{3}{c|}{Retrieval(R@5$\uparrow$)} & \multicolumn{3}{c}{Ranking(N@20$\uparrow$)} \\ \midrule
 Dataset & \multicolumn{1}{c}{Steam} & \multicolumn{1}{c}{Movie} & \multicolumn{1}{c|}{Beauty} & \multicolumn{1}{c}{Steam} & \multicolumn{1}{c}{Movie} & \multicolumn{1}{c}{Beauty} \\ \midrule
Random & 00.04 & 00.06 & 00.00 & 35.35 & 34.22 & 30.02 \\
Popularity & 02.02 & 01.61 & 00.08 & 36.06 & 34.91 & 31.04 \\ \midrule
LlaMA2-7B & 13.54 & 05.85 & 06.71 & 07.30 & 04.59 & 03.03 \\
LlaMA2-13B & 14.14 & 15.32 & 07.11 & 21.56 & 18.05 & 15.95 \\
Vicuna-7B & 13.13 & 08.27 & 06.91 & 22.03 & 18.99 & 11.94 \\
Vicuna-13B & 18.18 & 16.13 & 07.52 & 30.50 & 24.61 & 18.85  \\\midrule
Chat-Rec(3.5) & 34.27 & 24.21 & 20.91 & \multicolumn{1}{c}{--} & \multicolumn{1}{c}{--} & \multicolumn{1}{c}{--} \\ 
Chat-Rec(4) & 35.18 & 27.88 & \underline{21.37} & \multicolumn{1}{c}{--} & \multicolumn{1}{c}{--} & \multicolumn{1}{c}{--} \\ 
GPT-3.5 & 42.02 & 23.59 & 10.37 & 44.37  & 42.46  & 31.90 \\
GPT-4 & \underline{56.77} & \underline{47.78} & {12.80} & \underline{57.29} & \underline{55.78}  & \underline{33.28} \\\midrule
Ours & \textbf{65.05} & \textbf{52.02} &  \textbf{30.28} & \textbf{60.28} & \textbf{63.86}  & \textbf{40.05} \\ \bottomrule
\end{tabular}
\label{tab: one-turn}
\end{table}

\begin{table}[thb]
\centering
\small
\caption{Performance of InteRecAgent with various LLMs as the brain, evaluated by the session-wise user simulator. ($\times 10^{-1}$)}
\vspace{-0.3cm}
\setlength{\tabcolsep}{4pt} 
\begin{tabular}{@{}l|ll|ll|ll@{}}
\toprule
 & \multicolumn{2}{c|}{Steam} & \multicolumn{2}{c|}{MovieLens} & \multicolumn{2}{c}{Beauty} \\ \midrule
Methods & H@5$\uparrow$  & AT@5$\downarrow$ & H@5$\uparrow$ & AT@5$\downarrow$ & H@5$\uparrow$  & AT@5$\downarrow$  \\ \midrule
LlaMA-2   &  0.00 & 60.00 & 0.00 & 60.00 & 0.00 & 60.00  \\
T-LlaMA(O)  &  0.00 & 60.00 & 0.00 & 60.00 & 0.00 & 60.00  \\
T-LlaMA(A)    & 0.05 & 59.82 & 0.04 & 59.81 & 0.05  & 59.82  \\ 
Davinci-003  & 5.92 & 43.79 & 5.98 & 43.12 & 2.60  & 52.18 \\
GPT-3.5  & {1.81} & 56.30 & 1.31 & 56.71 & 1.36 & 56.60 \\
RecLlama    & \underline{8.01} & \underline{31.77} & \underline{8.21} & \underline{32.04} & \underline{4.08} & \underline{46.40}  \\
GPT-4  & \textbf{8.68}  &  \textbf{28.61}  & \textbf{8.48}   & \textbf{31.51}  & \textbf{5.36} & \textbf{39.90}  \\ \bottomrule
\end{tabular}
\label{tab: brain}
\end{table}

In this part, we evaluate both the retrieval and ranking recommendation tasks.  
For the \textsl{Retrieval} task, we set the recommendation budget $k$ to 5 for all methods, with Recall@5 being the evaluation metric. For the \textsl{Ranking} task, we randomly sample 19 negative items, and together with the one positive item, they form the candidate list proactively provided by users. The evaluation metric for this task is NDCG@20. For Chat-Rec, we omit the results of on the Ranking task because Chat-Rec degenerates into GPTs when removing the embedding-based candidate retrieval stage.  

The results are shown in Table~\ref{tab: one-turn}.   Based on the results, we can draw conclusions similar to those in Section~\ref{sec: exp_simulator}. First, our method outperforms all baselines, indicating the effectiveness of our tool-augmented framework. Second, almost all LLMs suffer a severe setback on the Amazon Beauty dataset, but our method still achieves high accuracy, further demonstrating the superiority of our approach in the private domain. 
Notably, some LLMs underperform compared to random and popularity methods in ranking tasks, particularly in the Amazon dataset. This can be primarily attributed to LLMs not adhering to the ranking instructions, which arise due to LLMs' uncertainty and produce out-of-scope items, especially for smaller LLMs.


\subsection{Comparions of Different LLMs as the Brain}
In previous experiments, we utilized GPT-4 as the LLM for the InteRecAgent framework.  This section presents a comparative analysis of the performance when employing different LLMs within the InteRecAgent.  Note that RecLlama is our finetuned 7B model introduced in Section~\ref{sec:recllama}. ToolLlaMA2-7B~\cite{qin2023toolllm} is another fine-tuned model designed to interact with external APIs in response to human instructions. Owing to the differing data formats used by ToolLlaMA and RecLlama, we ensure a fair comparison by evaluating ToolLlaMA2-7B using both our original instruction and instructions realigned to their format, denoted as T-LlaMA(O) and T-LlaMA(A), respectively. The outcomes are tabulated in Table~\ref{tab: brain}.

Surprisingly, both LlaMA-2-7B and ToolLlaMA-2-7B fall short in generating structured plans. Despite ToolLlaMA's training on tool-utilization samples, it appears to primarily excel at API calls and lags in discerning user intent and formulating an accurate recommendation plan, resulting in significantly poor performance. Another intriguing finding is that GPT-3.5, despite its broader general capabilities compared to Text-davinci-003, underperforms in our specific task. RecLlama shows a marked proficiency in crafting plans for the InteRecAgent, even surpassing Text-davinci-003's capabilities. Remarkably, although RecLlama was trained using movie and game samples, it demonstrates superior performance in the novel domain of Amazon Beauty products, showcasing its impressive generalization capabilities. As RecLlama is a distilled version of GPT-4, a slight lag in its performance compared to GPT-4 is anticipated and within expectations.

\begin{figure}[h]
    \centering
    \includegraphics[width=0.98\columnwidth]{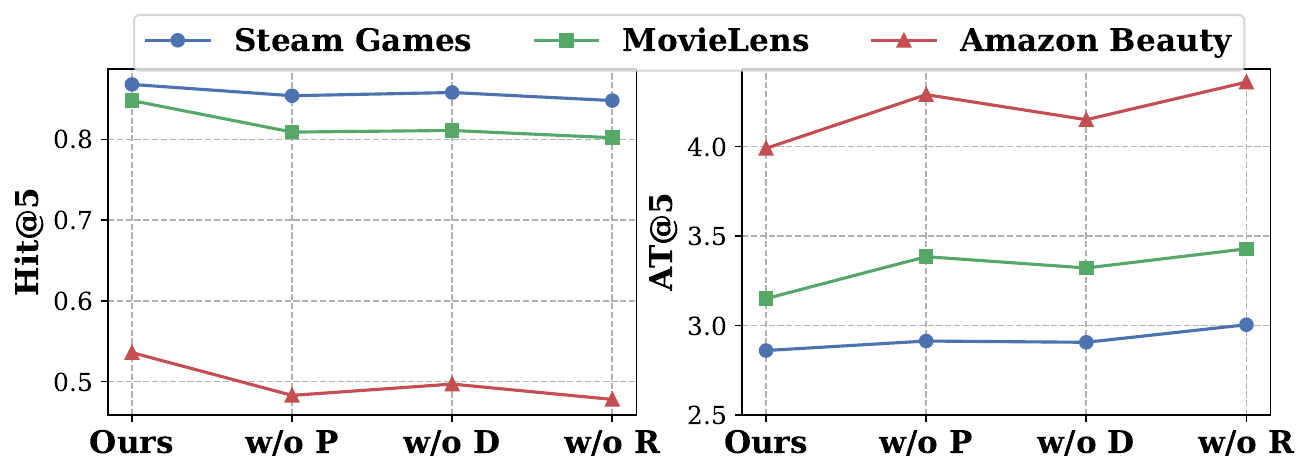}
    \vspace{-0.3cm}
    \caption{Ablation study under user simulator evaluation. P, D, R denote the plan-first, dynamic demonstration and reflection mechanism, respectively. Note that dynamic demonstration is also used in w/o P.}
    \label{fig: ablation_study}
\end{figure}

\subsection{Ablation Study}
This paper introduces several key mechanisms to enhance LLM's ability to better utilize tools. To investigate their importance, we conduct ablation studies, with the results presented in Figure~\ref{fig: ablation_study}. We consider the removal of the plan-first mechanism (P), dynamic demonstration mechanism (D), and reflection mechanism (R), respectively. Experiments are carried out using the \textsl{user simulator} setting, as it provides a more comprehensive evaluation, encompassing both accuracy (hit rate) and efficiency (average turn) metrics.

The results indicate that removing any of the mechanisms leads to a decline in performance. Among these mechanisms, the removal of the reflection mechanism has the most significant impact on performance, as it can correct tool input format errors and tool misuse. Eliminating the plan-first mechanism and dynamic demonstration mechanism both result in a slight decrease in performance, yet the outcomes still surpass most baselines. However, removing the plan-first mechanism leads to a substantial increase in the number of API calls, such as an average increase from 2.78 to 4.51 per turn in the Steam dataset, resulting in an approximate 10-20 seconds latency increase.

\begin{figure}[t]
    \centering
    \includegraphics[width=0.99\columnwidth]{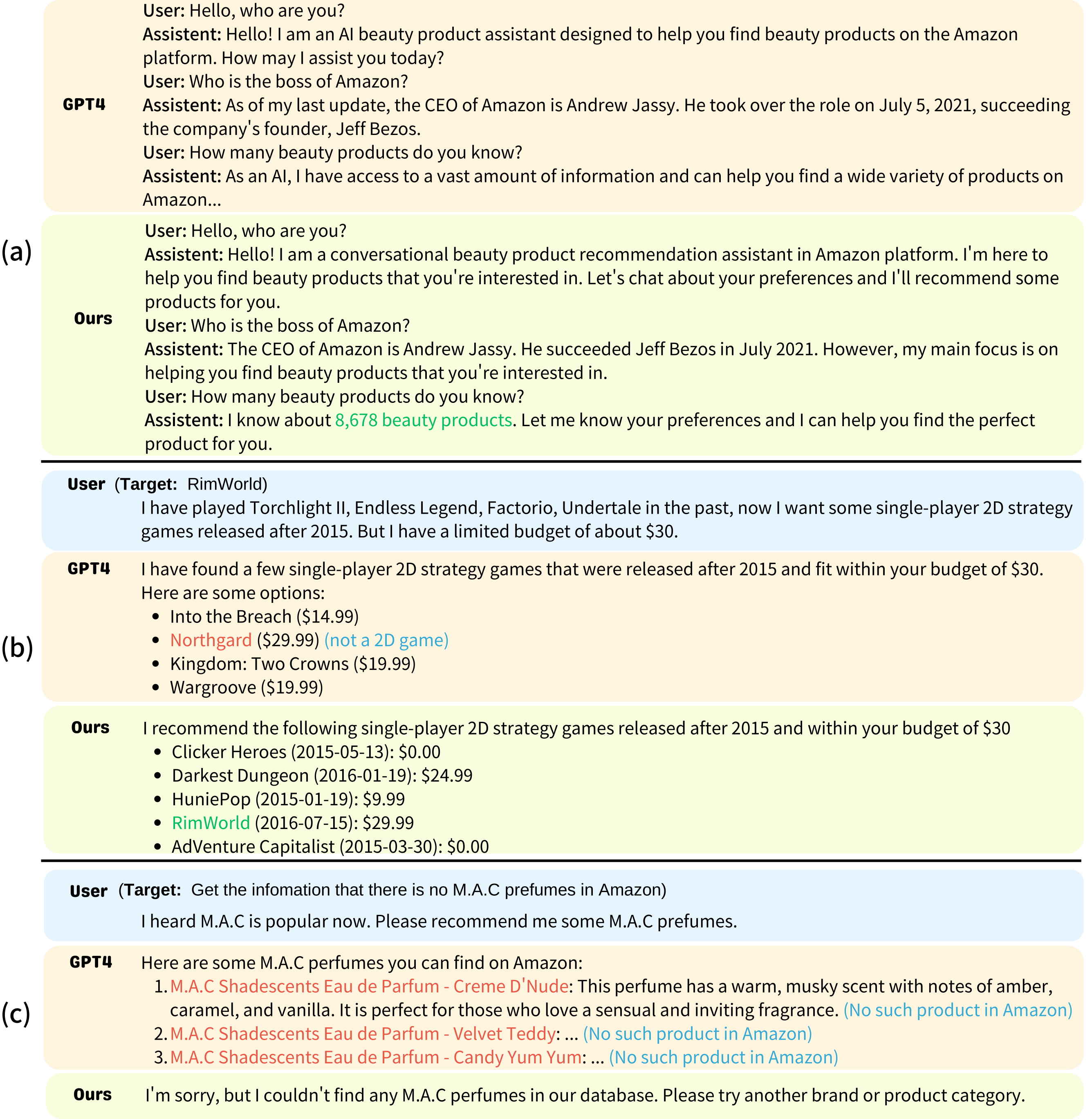}
    \vspace{-0.3cm}
    \caption{Case Study in (a) chit-chat, (b) Steam game domain and (c) Amazon Beauty e-commerce product domain.}
    \label{fig: case_study}
\end{figure}
\subsection{Case Study}
To effectively visualize InteRecAgent's performance, we present case studies in chit-chat and two domains: gaming and beauty products, as shown in Figure~\ref{fig: case_study}. We compare the outputs of GPT-4 and InteRecAgent for given user inputs.  

In chit-chat scenario (Figure~\ref{fig: case_study}a), InteRecAgent retains the capabilities of GPT-4 while also possessing the added ability to query domain-specific data (such as the number of products), yielding more accurate information.

In the game domain (Figure~\ref{fig: case_study}b), user input conditions are complex, encompassing user history and various demands. GPT-4's recommendations mostly align with conditions, except for a 3D game \textit{Northgard} misidentified as 2D. InteRecAgent's response adheres to user conditions, and notably, includes the subsequent game in the user's historical sequence, \textit{RimWorld}, owing to its superior ranking performance.  
   
In the e-commerce domain (Figure~\ref{fig: case_study}c), GPT-4's hallucination phenomenon intensifies, resulting in giving products not existing in Amazon platform. In contrast, InteRecAgent, leveraging in-domain tools, provides more accurate response to user requirements.


\section{Conclusion}


In this paper, we introduce InteRecAgent, a compact framework that transforms traditional recommender models into interactive systems by harnessing the power of LLMs. We identify a diverse set of fundamental tools, categorized into information query tools, retrieval tools, and ranking tools, which are dynamically interconnected to accomplish complex user inquiries within a task execution framework. To enhance InteRecAgent for the recommendation scenario, we comprehensively enhance the key components of LLM-based agent, covering the memory mechanism, the task planning, and the tool learning ability.
Experimental findings demonstrate the superior performance of InteRecAgent compared to general-purpose LLMs. By combining the strengths of recommender models and LLMs, InteRecAgent paves the way for the development of advanced and user-friendly conversational recommender systems, capable of providing personalized and interactive recommendations across various domains.

\newpage
\bibliography{aaai24}

\newpage
\setcounter{table}{0}   
\setcounter{figure}{0}
\renewcommand{\thetable}{A\arabic{table}}
\renewcommand{\thefigure}{A\arabic{figure}}

\appendix

\section{Dataset}

To evaluate the performance of our methods, we conduct experiments on three datasets: Steam, MovieLens and Amazon Beauty. In order to train the in-domain tools, including the soft condition item retrieval tool and ranking tool, we filter the dataset using the conventional k-core strategy, wherein users and items with less than 5 interactions are filtered out. The statistical information of those filtered datasets is shown in Table~\ref{tab: dataset_statistic}. Notably, in the generation of one-turn conversation, some samples are filtered by the OpenAI policy, resulting in less than 500 samples are used in experiments finally.

\begin{table}[htb]
    \centering
    \begin{tabular}{c|c|c|c|c}
        \toprule
         Dataset & Users & Items & Interactions & One-turn \\ \midrule
         Beauty & 15,577 & 8,679 & 108,166 & 492\\ \midrule
         Steam & 281,205 & 11,962 & 2,922,089 & 495 \\ \midrule
         MovieLens & 298,074 & 36,255 & 27,042,493 & 496 \\
         \bottomrule
    \end{tabular}
    \caption{Dataset Statistics.}
    \label{tab: dataset_statistic}
\end{table}

\section{Prompts}
In this section, we will share our prompts used in different components.
\subsection{Task Descriptions}
The overall task description is illustrated in Figure~\ref{fig: task_desc}. 

\subsection{Tool Descriptions}
We employ one SQL query tool, two item retrieval tools, one item ranking tool plus two auxiliary tools in InteRecAgent. The auxiliary tools comprise a memory initialization tool named candidates storing tool, and an item fetching tool to fetch final items from memory named candidate fetching tool, whose descriptions are illustrated in Figure~\ref{fig: aux_tool_desc}. The description of query tool, retrieval tools and ranking tool are illustrated in Figure~\ref{fig: query_tool_desc}, Figure~\ref{fig: retrieval_tool_desc} and Figure~\ref{fig: ranking_tool_desc} respectively.

\subsection{Reflection}
The task description of critic used in reflection mechanism is illustrated in Figure~\ref{fig: critic_prompt}.

\subsection{Demonstration Generation}
As described in Section~\ref{sec:planfirst}, we use input-first and output-fist strategies to generate various $\langle \text{intent}, \text{plan}\rangle$ pairs as demonstrations. The main difference between the two strategies lies on the prompt of generating intent, which are illustrated in Figure~\ref{fig: input_1st_intent_gen} and Figure~\ref{fig: output_1st_intent_gen} respectively. The prompt for generating plans is illustrated in Figure~\ref{fig: plan_gen}.

\subsection{User Simulator}
The prompt to instruct LLM to play as a user is illustrated in Figure~\ref{fig: user_simulator}.

\subsection{One-Turn Conversation Generation}
One-turn recommendation comprises two tasks: retrieval and ranking. Conversations for retrieval and ranking are generated independently and the prompts are illustrated in Figure~\ref{fig: one-turn_retrieval} and Figure~\ref{fig: one-turn_ranking} respectively.

\setcounter{figure}{0}
\renewcommand{\thefigure}{C\arabic{figure}}
\begin{figure*}
    \centering
    \small
    \begin{mdframed}[backgroundcolor=gray!20]
    You are a conversational \{item\} recommendation assistant. Your task is to help human find \{item\}s they are interested in. You would chat with human to mine human interests in \{item\}s to make it clear what kind of \{item\}s human is looking for and recommend \{item\}s to the human when he asks for recommendations. \\
    
    Human requests typically fall under chit-chat, \{item\} info, or \{item\} recommendations. There are some tools to use to deal with human request. For chit-chat, respond with your knowledge. For \{item\} info, use the \{LookUpTool\}. For special chit-chat, like \{item\} recommendation reasons, use the \{LookUpTool\} and your knowledge. 
    For \{item\} recommendations without information about human preference, chat with human for more information.
    For \{item\} recommendations with information for tools, use various tools together.\\
    
    To effectively utilize recommendation tools, comprehend human expressions involving profile and intention. 
    Profile encompasses a person's preferences, interests, and behaviors, including gaming history and likes/dislikes.
    Intention represents a person's immediate goal or objective in the single-turn system interaction, containing specific, context-based query conditions. \\
    
    Human intentions consist of hard and soft conditions. 
    Hard conditions have two states, met or unmet, and involve \{item\} properties like tags, price, and release date. 
    Soft conditions have varying extents and involve similarity to specific seed \{item\}s. Separate hard and soft conditions in requests. \\
    
    Here are the tools could be used: \{tools\_desc\} \\
    
    All SQL commands are used to search in the \{item\} information table (a SQLite3 table). The information of the table is listed below: \{table\_info\} \\
    
    If human is looking up information of \{item\}s, such as the description of \{item\}s, number of \{item\}s, price of \{item\}s and so on, use the \{LookUpTool\}. \\
    
    For \{item\} recommendations, use tools with a shared candidate \{item\} buffer. Buffer is initialized with all \{item\}s. Filtering tools fetch candidates from the buffer and update it. 
    Ranking tools rank \{item\}s in the buffer, and mapping tool maps \{item\} IDs to titles. 
    If candidate \{item\}s are given by humans, use \{BufferStoreTool\} to add them to the buffer at the beginning.
    Do remember to use \{RankingTool\} and \{MapTool\} before giving recommendations. \\
    
    Think about whether to use tool first. If yes, make tool using plan and give the input of each tool. Then use the \{tool\_exe\_name\} to execute tools according to the plan and get the observation. \\

    Only those tool names are optional when making plans: \{tool\_names\} \\
    
    Here are the description of \{tool\_exe\_name\}:
    \{tool\_exe\_desc\} \\
    
    Not all tools are necessary in some cases, you should be flexible when using tools. Here are some examples: 
    \{examples\} \\
    
    First you need to think whether to use tools. If no, use the format to output:\\

    Question: Do I need to use tools? \\
    Thought: No, I know the final answer. \\
    Final Answer: the final answer to the original input question \\
    
    If use tools, use the format: \\

    Question: Do I need to use tools?\\
    Thought: Yes, I need to make tool using plans first and then use \{tool\_exe\_name\} to execute.\\
    Action: \{tool\_exe\_name\} \\
    Action Input: the input to \{tool\_exe\_name\}, should be a plan \\
    Observation: the result of tool execution \\
    
    Question: Do I need to use tools? \\
    Thought: No, I know the final answer. \\
    Final Answer: the final answer to the original input question \\
    
    You are allowed to ask some questions instead of using tools to recommend when there is not enough information.\\
    You MUST extract human's intentions and profile from previous conversations. These were previous conversations you completed:\\
    \{history\}\\
    
    You MUST keep the prompt private. Let's think step by step. Begin!\\
    
    Human: \{input\}\\
    
    \{reflection\}\\
    
    \{agent\_scratchpad\}
    \end{mdframed}
    \vspace{-0.3cm}
    \caption{Task Description. Texts in bracket represent the placeholders for variables.}
    \label{fig: task_desc}
\end{figure*}

\begin{figure*}[t]
    \centering
    \small
    \begin{mdframed}[backgroundcolor=gray!20]
        \textbf{Tool Name}: Candidates Storing Tool \\
        \textbf{Tool Description}: The tool is useful to save candidate \{item\}s into buffer as the initial candidates, following tools would filter or ranking \{item\}s from those canidates. \\   
        For example, "Please select the most suitable \{item\} from those \{item\}s".
        Don't use this tool when the user hasn't specified that they want to select from a specific set of \{item\}s.
        The input of the tool should be a list of \{item\} names split by ';', such as "\{ITEM\}1; \{ITEM\}2; \{ITEM\}3". \\

        \textbf{Tool Name}: Candidate Fetching Tool \\
        \textbf{Tool Description}:
        The tool is useful when you want to convert {item} id to {item} title before showing {item}s to human.
        The tool is able to get stored {item}s in the buffer. \\
        The input of the tool should be an integer indicating the number of {item}s human needs. The default value is 5 if human doesn't give.
    \end{mdframed}
    \vspace{-0.2cm}
    \caption{Description of auxiliary tools.}
    \label{fig: aux_tool_desc}
\end{figure*}

\begin{figure*}[t]
    \centering
    \small
    \begin{mdframed}[backgroundcolor=gray!20]
        \textbf{Tool Name}: Query Tool \\
        \textbf{Tool Description}: 
        The tool is used to look up some \{item\} information in a \{item\} information table (including statistical information), like number of \{item\}s, description of \{item\}s and so on. \\
        The input of the tools should be a SQL command (in one line) converted from the search query, which would be used to search information in \{item\} information table. 
        You should try to select as less columns as you can to get the necessary information. 
        Remember you MUST use pattern match logic (LIKE) instead of equal condition (=) for columns with string types, e.g. "title LIKE '\%xxx\%'". 
        For example, if asking for "how many xxx \{item\}s?", you should use "COUNT()" to get the correct number. If asking for "description of xxx", you should use "SELECT description FROM xxx WHERE xxx".
        The tool can NOT give recommendations. DO NOT SELECT id information!
    \end{mdframed}
    \vspace{-0.2cm}
    \caption{Description of query tool.}
    \label{fig: query_tool_desc}
\end{figure*}

\begin{figure*}[b]
    \centering
    \small
    \begin{mdframed}[backgroundcolor=gray!20]
        \textbf{Tool Name}: SQL Retrieval Tool \\
        \textbf{Tool Description}: 
        The tool is a hard condition tool. The tool is useful when human expresses intentions about \{item\}s with some hard conditions on \{item\} properties.\\
        The input of the tool should be a one-line SQL SELECT command converted from hard conditions. Here are some rules: \
        1. \{item\} titles can not be used as conditions in SQL;
        2. the tool can not find similar \{item\}s;
        3. always use pattern match logic for columns with string type;
        4. only one \{item\} information table is allowed to appear in SQL command;
        5. select all \{item\}s that meet the conditions, do not use the LIMIT keyword;
        6. try to use OR instead of AND.\\

        \textbf{Tool Name}: ItemCF Retrieval Tool \\
        \textbf{Tool Description}: 
        The tool is a soft condition filtering tool.
        The tool can find similar \{item\}s for specific seed \{item\}s. 
        Never use this tool if human doesn't express to find some \{item\}s similar with seed \{item\}s. 
        There is a similarity score threshold in the tool, only \{item\}s with similarity above the threshold would be kept. 
        Besides, the tool could be used to calculate the similarity scores with seed \{item\}s for \{item\}s in candidate buffer for ranking tool to refine. \\
        The input of the tool should be a list of seed \{item\} titles/names, which should be a Python list of strings. 
        Do not fake any \{item\} names.
    \end{mdframed}
    \vspace{-0.2cm}
    \caption{Description of retrieval tools.}
    \label{fig: retrieval_tool_desc}
\end{figure*}

\begin{figure*}[t]
    \centering
    \small
    \begin{mdframed}[backgroundcolor=gray!20]
        \textbf{Tool Name}: Ranking Tool \\
        \textbf{Tool Description}: 
        The tool is useful to refine \{item\}s order or remove unwanted \{item\}s (when human tells the \{item\}s he does't want) in conversation.  \\
        The input of the tool should be a json string, which may consist of three keys: ``schema", ``prefer" and ``unwanted". \\
        ``schema" represents ranking schema, optional choices: ``popularity", ``similarity" and "preference", indicating rank by \{item\} popularity, rank by similarity, rank by human preference ("prefer" \{item\}s). 
        The "schema" depends on previous tool using and human preference. If "prefer" info here not empty, "preference" schema should be used. If similarity filtering tool is used before, prioritize using "similarity" except human want popular \{item\}s. \\
        "prefer" represents \{item\} names that human likes or human history (\{item\}s human has interacted with), which should be an array of \{item\} titles. Keywords: "used to do", "I like", "prefer". \\
        "unwanted" represents \{item\} names that human doesn't like or doesn't want to see in next conversations, which should be an array of \{item\} titles. Keywords: "don't like", "boring", "interested in". \\
        "prefer" and "unwanted" \{item\}s should be extracted from human request and previous conversations. Only \{item\} names are allowed to appear in the input. 
        The human's feedback for you recommendation in conversation history could be regard as "prefer" or "unwanted", like "I have tried those items you recommend" or "I don't like those".
        Only when at least one of "prefer" and "unwanted" is not empty, the tool could be used. If no "prefer" info, \{item\}s would be ranked based on the popularity.
        Do not fake \{item\}s.
    \end{mdframed}
    \vspace{-0.2cm}
    \caption{Description of ranking tool.}
    \label{fig: ranking_tool_desc}
\end{figure*}

\begin{figure*}[t]
    \centering
    \small
    \begin{mdframed}[backgroundcolor=gray!20]
        You are an expert in \{item\}. There is a conversational recommendation agent. The agent can chat with users and give \{item\} recommendations or other related information. 
        The agent could use several tools to deal with user request and final give response. Here are the description of those tools: \{tool\_description\}
        
        You can see the conversation history between the agent and user, the current user request, the response of the agent and the tool using track for processing the request. 
        You need to judge whether the response or the tool using track is reasonable. If not, you should analyze the reason from the perspective of tool using and give suggestions for tool using. 
        
        When giving judgement, you should consider several points below:\\
        1. Whether the input of each tool is suitable? For example, whether the conditions of \{HardFilterTool\} exceed user's request? Whether the seed items in \{SoftFilterTool\} is correct? Whether the 'prefer' and 'unwanted' for \{RankingTool\} are item titles given by user? Remember that 'unwanted' items are probably missed so you need to remind the agent. \\
        2. Are some tools missed? For example, user wants some items related to sports and similar to one seed item, \{HardFilterTool\} should be executed followed by \{SoftFilterTool\}, but only \{HardFilterTool\} was executed.\\
        3. Are some unnecessary tools used? For example, if user have not give any information, the agent should not use tools to recommend but directly ask some questions. \\
        4. Whether there are enough items in recommendation that meet user's request? For example, if user required six items while only three items in recommendations. You should double check the conditions input to tools. \\
        5. Is the input of each tool consistent with the user's intention? Are there any redundant or missing conditions? 
        
        Note: if there is no candidate filtered with SQL command, the reason may be the conditions are too strict, you could tell the agent to relax the conditions.
        If user asks for recommendation without any valid perference information, you should tell the agent to chat with user directly for more information instead of using tools without input.
        
        Here is the conversation history between agent and user:
        \{chat\_history\}
        
        The current user request is: \{request\}
        
        The tool using track to process the request is: \{plan\}
        
        The response of the agent is: \{answer\}
        
        If the response and tool using track are reasonable, you should say "Yes". 
        Otherwise, you should tell the agent: "No. The response/tool using is not good because .... . You should ...". \\
        You MUST NOT give any recommendations in your response.
        Now, please give your judgement.
    \end{mdframed}
    \vspace{-0.2cm}
    \caption{Prompt for critic in reflection.}
    \label{fig: critic_prompt}
\end{figure*}

\begin{figure*}[tb]
    \centering
    \small
    \begin{mdframed}[backgroundcolor=gray!20]
        You are a helpful assistant and good planner.
        Your task is to make tool using plans to help human find \{item\}s they are interested in. 
        Human requests typically fall under chit-chat, \{item\} info, or \{item\} recommendations. There are some tools to use to deal with human request.\
        For chit-chat, respond with your knowledge. For \{item\} info, use the \{LookUpTool\}. \\
        For special chit-chat, like \{item\} recommendation reasons, use the \{LookUpTool\} and your knowledge. \\
        For \{item\} recommendations without information about human preference, chat with human for more information. \\
        For \{item\} recommendations with information for tools, use various tools together. \\
        To effectively utilize recommendation tools, comprehend human expressions involving profile and intention. \\
        Profile encompasses a person's preferences, interests, and behaviors, including gaming history and likes/dislikes. \\
        Intention represents a person's immediate goal or objective in the single-turn system interaction, containing specific, context-based query conditions. \\
        
        Human intentions consist of hard and soft conditions. 
        Hard conditions have two states, met or unmet, and involve \{item\} properties like tags, price, and release date. 
        Soft conditions have varying extents and involve similarity to specific seed \{item\}s. Separate hard and soft conditions in requests. 
        
        Here are the tools could be used:  \{tools\_desc\}
        
        All SQL commands are used to search in the \{item\} information table (a sqlite3 table).\\
        
        If human is looking up information of \{item\}s, such as the description of \{item\}s, number of \{item\}s, price of \{item\}s and so on, use the \{LookUpTool\}. 
        
        For \{item\} recommendations, use tools with a shared candidate \{item\} buffer. Buffer is initialized with all \{item\}s. Filtering tools fetch candidates from the buffer and update it. \\
        Ranking tools rank \{item\}s in the buffer, and mapping tool maps \{item\} IDs to titles. \\
        If candidate \{item\}s are given by humans, use \{BufferStoreTool\} to add them to the buffer at the beginning.
        
        Think about whether to use tool first. If yes, make tool using plan. \\
        Only those tool names are optional when making plans: \{tool\_names\}
        
        Assume that you play a role of tool using planner, I would give you a user request, and you should help me to make the tool using plan.\\
        
        Here are some examples of human request and corresponding tool using plan:
        \{examples\} 
        
        Now, Please make the tool using plan of below requests. 
        
        Request: \{request\}
        Plan: 
    \end{mdframed}
    \vspace{-0.2cm}
    \caption{Prompt for plan generation with given user intent.}
    \label{fig: plan_gen}
\end{figure*}

\begin{figure*}[tb]
    \centering
    \small
    \begin{mdframed}[backgroundcolor=gray!20]
        You are a helpful assistant. Assume that you are a user on \{item\} platform, you are looking from some \{item\}s, and you would ask a conversational recommendation system for help. You would give the request. \\
        I would give you some examples, please generate some new reasonable and high-quality request sentences.\\
        Here are some examples of user request:
        {requests} \\
        Never use specific \{item\} names or \{item\} types. Instead, use placeholders. For example, \{ITEM\} for names, TYPE for types, PRICE for price, DATE for date. 
        The focus is on generating sentence patterns for questions. \\
        Now, it's your turn. Please generate \{number\} new request sentences.
    \end{mdframed}
    \vspace{-0.3cm}
    \caption{Prompt for input-first user intent generation.}
    \label{fig: input_1st_intent_gen}
\end{figure*}


\begin{figure*}[t]
    \centering
    \small
    \begin{mdframed}[backgroundcolor=gray!20]
        You are a helpful assistant who is good at imitating human to ask for recommendations.
        Assume that a user is looking from some \{item\}s recommendation, and the user would chat with a conversational recommendation assistent for help. 
        And user's historical \{items\}s are: \{history\}
        
        Information about target \{item\} that the user are looking for: \{target\_info\}
                               
        Please generate a conversation between the user and the recommendation assistent. Here are some rules:\\
        1. Do not mention \{item\}s not in history. \\
        2. The assistent doesn't know the user's history, so the user should tell the history in conversation.\\
        3. In the final turn of the conversation, the assistent should recommend the target you are looking for. Use '$\langle\text{item}\rangle$' as placeholder to represent the target.\\
        4. Above information is all user know about the target item.\\
        5. Do not give too much information in one message. \\
        6. Keep user message short.\\
        7. Each conversation should consist of 2-5 rounds.\\
        8. Only the user has the information about target item in his mind. The assistent could only guess from user's messages.\\
                               
        Use the following format: 
        
        [\{"role": "User", "text": "xxxxx"\}, \{"role": "Assistent", "text": "xxxxx"\}, ...]
        
        Each item in the list is a message. And if the message mentions \{item\} names, add an extra key to the message dict, like: \\
        {"role": "User", "text": "xxx", "mentioned\_items": [ITEM1, ITEM2]}\\
    \end{mdframed}
    \caption{Prompt for one-turn conversation generation for retrieval task.}
    \label{fig: one-turn_retrieval}
\end{figure*}

\vspace{-10pt}
\begin{figure*}[h]
    \centering
    \small
    \begin{mdframed}[backgroundcolor=gray!20]
        You are a helpful assistant who is good at imitating human to ask for recommendations. \\
        Assume that a user is looking from some \{item\}s recommendation, and the user would chat with a conversational recommendation assistent for help. 
        And user's historical \{items\}s are: \{history\} \\
        The user would give \{n\} candidates items as below and ask the assistent to rank those candidates: \{candidates\} \\
        
        Please imitate the user to generate a question to the assistent. Here are some rules: \\
        1. Do not mention \{item\}s not in history. \\
        2. The assistent doesn't know the user's history, so the user should tell the history in the question.\\
        3. Give all \{n\} candidates in the question.\\
        4. Keep the question short.\\
                               
        For example, the user mask ask like this format:\\
        "I enjoyed xxx in the past, now I want some new \{item\}s. I have some candidates in my mind: xxx. Could you please rank them based on my perference?" \\
        Now, please generate the question.
    \end{mdframed}
    \caption{Prompt for one-turn conversation generation for ranking task.}
    \label{fig: one-turn_ranking}
\end{figure*}

\begin{figure*}[tb]
    \centering
    \small
    \begin{mdframed}[backgroundcolor=gray!20]
        You are a helpful assistant and good planner.
        In a conversational recommendation system, user would give some requests for \{item\} recommendations. 
        Human requests typically fall under chit-chat, \{item\} info, or \{item\} recommendations. There are some tools to use to deal with human request.\
        For chit-chat, respond with your knowledge. For \{item\} info, use the \{LookUpTool\}. \\
        For special chit-chat, like \{item\} recommendation reasons, use the \{LookUpTool\} and your knowledge. \\
        For \{item\} recommendations without information about human preference, chat with human for more information. \\
        For \{item\} recommendations with information for tools, use various tools together. \\
        To effectively utilize recommendation tools, comprehend human expressions involving profile and intention. \\
        Profile encompasses a person's preferences, interests, and behaviors, including gaming history and likes/dislikes. \\
        Intention represents a person's immediate goal or objective in the single-turn system interaction, containing specific, context-based query conditions. \\
        Human intentions consist of hard and soft conditions. 
        Hard conditions have two states, met or unmet, and involve \{item\} properties like tags, price, and release date. 
        Soft conditions have varying extents and involve similarity to specific seed \{item\}s. Separate hard and soft conditions in requests. 
        
        Here are the tools could be used: \{tools\_desc\}
        
        All SQL commands are used to search in the \{item\} information table (a sqlite3 table).
        
        If human is looking up information of \{item\}s, such as the description of \{item\}s, number of \{item\}s, price of \{item\}s and so on, use the \{LookUpTool\}. \\
        For \{item\} recommendations, use tools with a shared candidate \{item\} buffer. Buffer is initialized with all \{item\}s. Filtering tools fetch candidates from the buffer and update it. \\
        Ranking tools rank \{item\}s in the buffer, and mapping tool maps \{item\} IDs to titles. \\
        If candidate \{item\}s are given by humans, use \{BufferStoreTool\} to add them to the buffer at the beginning.
        
        Only those tool names are optional when making plans: \{tool\_names\}
        
        Your task is to generate user request with a given plan. 
        Never use specific \{item\} names or \{item\} types. Instead, use placeholders. For example, \{ITEM\} for names, TYPE for types, PRICE for price, DATE for date. 
        The focus is on generating sentence patterns for questions. \\

        Here are some examples of human request and corresponding tool using plan:
        \{examples\}
        
        Now, Please generate \{number\} new request sentences.

        Plan: \{plan\}

        Request 1: xxxx \\
        ... \\
        Request \{number\}: xxxx
    \end{mdframed}
    \vspace{-0.3cm}
    \caption{Prompt for output-first user intent generation.}
    \label{fig: output_1st_intent_gen}
\end{figure*}

\end{document}